\documentclass[superscriptaddress,twocolumn,aps,showpacs,pra,preprintnumbers]{revtex4-1}

\usepackage{soul}
\usepackage{textcomp}
\usepackage{graphicx}
\usepackage{amsmath}
\usepackage{xcolor}
\usepackage[hidelinks]{hyperref}

\newcommand{\be}{\begin{equation}}
\newcommand{\ee}{\end{equation}}

\begin{document}

\title{Dysprosium dipolar Bose-Einstein condensate with broad Feshbach resonances}

\author{E. Lucioni}\email{lucioni@lens.unifi.it}
\affiliation{LENS and Dip. di Fisica e Astronomia, Universit$\grave{\rm a}$ di Firenze, 50019 Sesto Fiorentino, Italy}
\affiliation{CNR-INO, S.S. A.~Gozzini di Pisa, via Moruzzi 1, 56124 Pisa, Italy}
\author{L. Tanzi}
\affiliation{LENS and Dip. di Fisica e Astronomia, Universit$\grave{\rm a}$ di Firenze, 50019 Sesto Fiorentino, Italy}
\affiliation{CNR-INO, S.S. A.~Gozzini di Pisa, via Moruzzi 1, 56124 Pisa, Italy}
\author{A. Fregosi}
\affiliation{Dipartimento di Fisica, Universit$\grave{\rm a}$ di Pisa, Largo B. Pontecorvo, 56127 Pisa, Italy}
\affiliation{CNR-INO, S.S. A.~Gozzini di Pisa, via Moruzzi 1, 56124 Pisa, Italy}
\author{J. Catani}
\affiliation{LENS and Dip. di Fisica e Astronomia, Universit$\grave{\rm a}$ di Firenze, 50019 Sesto Fiorentino, Italy}
\affiliation{CNR-INO, S.S. Sesto Fiorentino, 50019 Sesto Fiorentino, Italy}
\author{S. Gozzini}
\affiliation{CNR-INO, S.S. A.~Gozzini di Pisa, via Moruzzi 1, 56124 Pisa, Italy}
\author{M. Inguscio}
\affiliation{LENS and Dip. di Fisica e Astronomia, Universit$\grave{\rm a}$ di Firenze, 50019 Sesto Fiorentino, Italy}
\affiliation{CNR-INO, S.S. Sesto Fiorentino, 50019 Sesto Fiorentino, Italy}
\author{A. Fioretti}
\affiliation{CNR-INO, S.S. A.~Gozzini di Pisa, via Moruzzi 1, 56124 Pisa, Italy}
\author{C. Gabbanini}
\affiliation{CNR-INO, S.S. A.~Gozzini di Pisa, via Moruzzi 1, 56124 Pisa, Italy}
\author{G. Modugno}
\affiliation{LENS and Dip. di Fisica e Astronomia, Universit$\grave{\rm a}$ di Firenze, 50019 Sesto Fiorentino, Italy}
\affiliation{CNR-INO, S.S. A.~Gozzini di Pisa, via Moruzzi 1, 56124 Pisa, Italy}

\date{\today}

\begin{abstract}
We produce Bose-Einstein condensates of $^{162}$Dy atoms employing an innovative technique based on a resonator-enhanced optical trap that allows efficient loading from the magneto-optical trap. We characterize the scattering properties of the ultracold atoms for magnetic fields between 6 and 30~G. In addition to the typical chaotic distribution of narrow Feshbach resonances in Lanthanides, we discover two rather isolated broad features at around 22~G and 27~G. A characterization using the complementary measurements of losses, thermalization, anisotropic expansion and molecular binding energy points towards resonances of predominant s-wave character. Such resonances will ease the investigation of quantum phenomena relying on the interplay between dipole and contact interactions.
\end{abstract}


\maketitle

Dipolar atomic Bose-Einstein condensates (dBEC) are proving to be excellent platforms for the study of a range of quantum phenomena relying on the interplay between the anisotropic long-range dipole-dipole interaction and the isotropic contact one. Recent experiments with dBEC demonstrated the existence of an unexpected quantum liquid phase, emerging for attractive mean-field interactions and stabilized by quantum fluctuations \cite{Kadau2016,FerrierPRL,Chomaz2016,Schmitt2016}, showed the possibility to study lattice physics beyond the standard Bose-Hubbard model \cite{Baier2016}, and revealed first signatures of peculiar roton excitations \cite{Chomaz2018} and scissors oscillations \cite{Stringari}. All these observations rely on the large magnetic moment available in Lanthanides, and require a fine control of the relative strength of dipolar and contact interactions. However, so far only very narrow Feshbach resonances, with widths of the order of tens of mG, have been employed to this scope. In fact, the complex electronic structure of such atoms, responsible for their large magnetic dipole moment, also leads to a strong anisotropy of the van der Waals interaction, which gives rise to an extremely dense chaotic distribution of narrow Feshbach resonances \cite{Frisch2014,Maier2015}. Dysprosium is the most magnetic atom available, whose magnetic dipole moment of 9.93~$\mu_B$ results in a dipolar length $a_{dd}\simeq$~130~$a_0$. In the ground state of $^{164}$Dy, besides the chaotic spectrum, two very broad Feshbach resonances with $\Delta\simeq$~30~G have been observed and characterized \cite{PfauJulienne}. Their practical use is however questionable, since the resonances poles are surrounded by many narrow resonances. The other bosonic isotope, $^{162}$Dy, has been characterized only up to 6~G: only narrow resonances appear, with the largest width around 25 mG \cite{Baumann2014,Tang2016}.

In this work we report on the production of a dBEC of $^{162}$Dy and on the exploration of the resonance spectrum up to 30 G. The BEC is produced employing a large-volume optical trap enhanced by an in-vacuum optical resonator, which allows an efficient capture of atoms from the magneto-optical trap (MOT) using a low-power single-mode laser. This technique was so far used only with alkalis and Yb atoms \cite{resonatorLi, resonatorYb}; the application to dipolar Lanthanides is particularly interesting, because of their low optical polarizability \cite{Dzuba2011,Li2017,Ravensbergen2018}. We then employ ultracold samples at temperatures just above condensation to investigate the spectrum of Feshbach resonances. We discover two relatively isolated resonances with widths $\Delta\simeq$ 0.1-1~G comparable to the typical spacing between narrow resonances. Such resonances appear particularly appealing for a precise tuning of the contact interaction over a broad range, a possibility that was so far absent in highly magnetic atoms. Using complementary measurements of losses, thermalization, anisotropic expansion and molecular binding energy, we provide our best characterization of the resonances parameters. An analysis of the resonances that assumes a predominant s-wave character gives resonance strengths $s_{res}\simeq$~0.5.

We start by describing the experimental sequence employed to reach condensation. An atomic beam exits an effusive cell where a solid Dy sample is heated at 1110$^\circ$~C. The beam is collimated by a hot tube inside the cell, a cold skimmer outside the cell and a transverse cooling stage working on the broadest Dy transition at 421~nm ($\Gamma$/$2 \pi $=32~MHz, with $\Gamma$ the transition linewidth).
The atoms are slowed down from an initial average velocity of approximately 450~m/s to a velocity of a few m/s in a spin-flip Zeeman slower, also operating on the 421~nm transition. They are then caught in the MOT operating on the narrower transition at 626~nm ($\Gamma$/$2 \pi $=135~kHz) \cite{Lucioni2017}. The capture velocity of the MOT is artificially increased by a modulation of the laser frequency. By operating the MOT at large detuning ($\approx -35\, \Gamma$) gravity shifts the atomic cloud below the quadrupole center and the atoms get spontaneously polarized in the stretched Zeeman state ($m_J$=-8), as already observed in related setups \cite{Dreon2017}. We load the MOT during 7 seconds, then we perform a compression in order to increase the phase space density: the frequency broadening is switched off, the power of the MOT beams is reduced to 0.3~$I_{sat}$ and the laser frequency is set closer to the atomic resonance ($\approx -8~\Gamma$) \cite{Lucioni2017}. After the compression, the typical atom number in the MOT is $6 \times 10^7$, with a Gaussian RMS width of $450\,\mu$m in the horizontal plane and $150\,\mu$m along the vertical direction, at a typical temperature of $15\,\mu$K.

\begin{figure}[t]
\includegraphics[width=\columnwidth]{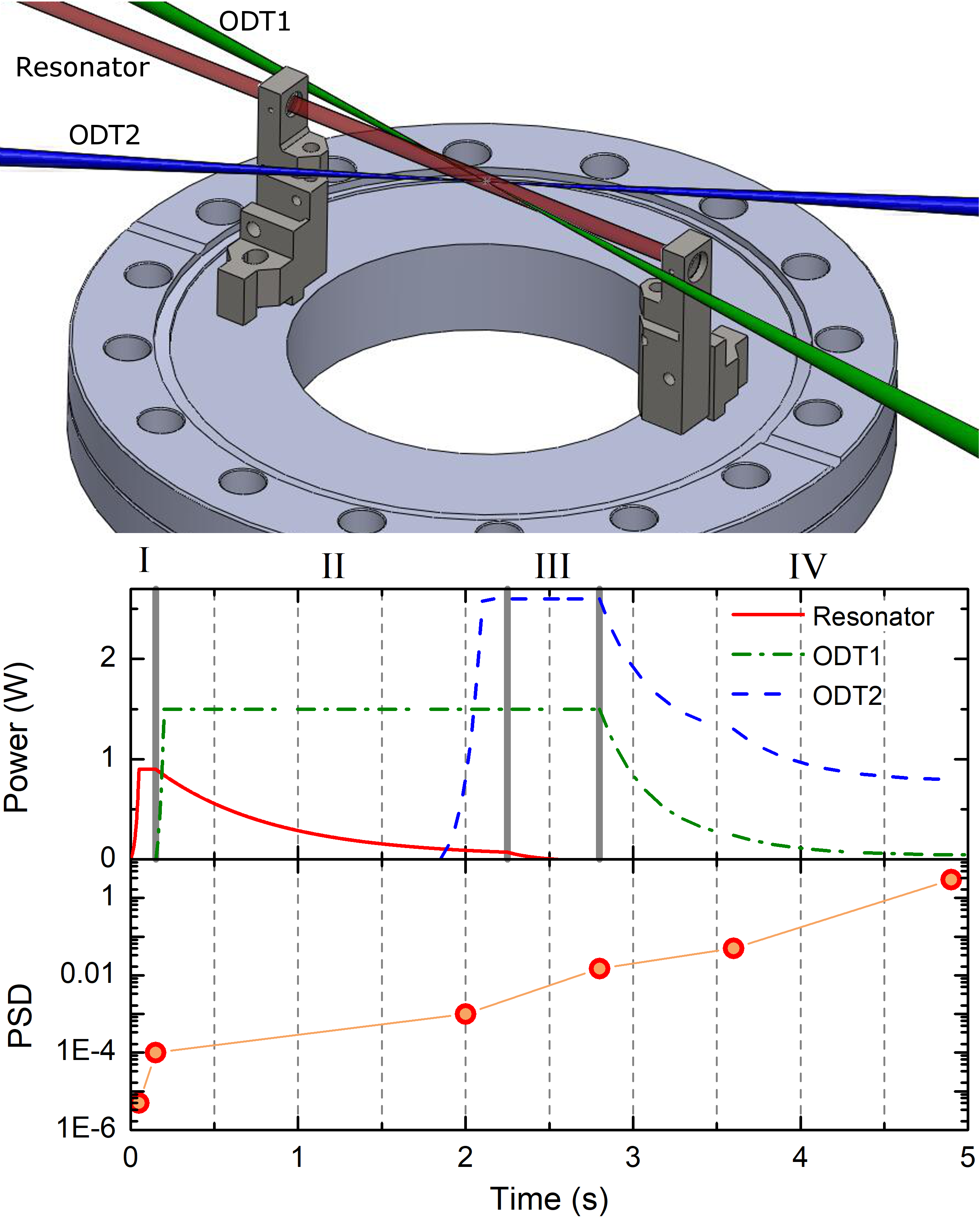}
\caption{Top: Schematic of the trapping potentials in the vacuum chamber: the resonator trap is shown in red, the optical traps ODT1 and ODT2 in green and in blue, respectively. The angle between ODT1 and the resonator is 8\textdegree, the angle between ODT1 and ODT2 is 40\textdegree. Bottom: Power of employed laser beams and phase space density (PSD) through the experimental cycle. The scheme is divided in stages: (I) MOT compression and resonator trap loading, (II) evaporation in the resonator, (III) transfer from the resonator to the crossed ODT1 and ODT2), and (IV) forced evaporation in the ODTs to Bose-Einstein condensation.}  \label{fig1}
\end{figure}

The primary optical trap is realized by the standing-wave pattern inside an in-vacuum optical resonator, seeded by a single-mode Nd:YAG laser at 1064~nm. With such a scheme we achieve large trapping volumes and trap depths without employing high-power multimode lasers, which tend to cause unwanted heating and losses \cite{multi_losses}. The resonator cavity is made up by two spherical mirrors with large curvature radius (3~m) at a reciprocal distance of 9~cm, with a measured finesse F=1050(20). By coupling 0.9 W of light into the cavity we obtain a trap depth of 200~$\mu$K with a waist of 320~$\mu$m, by using the scalar polarizability of 184.4~a.u. recently measured in \cite{Ravensbergen2018}. The light is actively frequency locked to the cavity by a fast feedback on the laser piezo and a slow feedback on the temperature of the laser crystal.

We ramp the power of the resonator trap up during the last part of the compression phase, when also the incoming atomic beam is blocked by a pneumatic shutter (stage I in Fig.\ref{fig1}). The geometrical superposition of the MOT with the trap is optimized by adjusting the compression parameters and the position of the MOT by means of small magnetic bias fields. The trap volume is comparable to the volume of the atomic cloud and therefore we load approximately half of the atoms of the compressed MOT, 3$\times 10^7$, at a temperature of $30~\mu$K. Such large loading efficiency allows us to operate with a MOT with smaller phase-space density than other setups \cite{Dreon2017,Frisch2012,MaierMOT,TangBEC}. At this stage, the atom number per lattice site is approximately 15000 and the trap frequencies are 105~Hz and 140~kHz in the radial and lattice direction, respectively. A uniform magnetic field of 3.335(3)~G is adiabatically switched on along the vertical direction in order to keep the atoms polarized in the $m_J$=-8 state.

We observe a strong dependence of the trap loading efficiency on the light polarization with respect to the dipoles orientation. In particular, for light polarization parallel to the dipoles we observe a light shift of the 626~nm transition of several $\Gamma$, which reduces the laser cooling efficiency in the presence of the trap, resulting in a poor loading efficiency. This observation suggest an anisotropic tensor part of the dynamical polarizability of the excited state. This effect has been recently studied in \cite{Chalopin2018} and, for Er, in \cite{Becher2017}. For the aim of the present work, we empirically adjust the polarization angle in order to optimize the loading efficiency: the best condition is when the light polarization is almost perpendicular to the dipoles.

Before starting the forced evaporation in the resonator trap, we ramp up a single beam optical trap (ODT1) with an angle of 8\textdegree with respect to the resonator. This beam has a waist of 41~$\mu$m and a power of 1.5~W. We then exponentially ramp the resonator trap depth down in 2100~ms until the vertical confinement is one tenth of the initial one. During the evaporation, the cold atoms get collected in the potential well created by ODT1 but cannot move along the longitudinal direction because of the lattice potential (stage II in Fig.\ref{fig1}). Afterwards we ramp up a second beam (ODT2) with an angle of 40\textdegree with respect to ODT1. This beam is elliptically shaped with a horizontal (vertical) waist of 81~$\mu$m (36~$\mu$m) and has a power of 2.6~W. At this point the resonator power is further ramped down to 10$^{-4}$ of the initial power allowing the atoms to collect in the crossed region between ODT1 and ODT2 (stage III in Fig.\ref{fig1}). Power is not set to zero to preserve the active frequency locking, however the residual lattice potential due to the resonator is smaller than 0.1 recoil energy. In the crossed trap we typically have 10$^6$ atoms at a temperature of 4~$\mu$K. The trap frequencies are $(\nu_v,\nu_{h_1},\nu_{h_2})=(400,300,80)$~Hz, with $v$ $(h_1,h_2)$ denoting the vertical (horizontal) direction. We perform evaporative cooling by reducing the trap powers with exponential ramps (stage IV in Fig.\ref{fig1}). The ramps are shaped in such a way that evaporation mainly occurs along the vertical direction. During the last phase of the evaporation we pay particular attention in keeping the ratio between the vertical trap frequency and the average on plane trap frequency larger than 3, in order to allow the BEC formation avoiding dipolar collapse \cite{Koch2008}; this is possible thanks to the elliptical shape chosen for ODT2. In order to have a pure BEC with negligible thermal component, the power of ODT1 (ODT2) is reduced to 50~mW (800~mW). The final trap frequencies are $(\nu_v,\nu_{h_1},\nu_{h_2})=(140,80,30)$~Hz. We typically produce BECs of $4\times 10^4$ atoms with transition temperature around 80~nK. The full experimental sequence lasts 13 seconds.

\begin{figure}[t]
\includegraphics[width=\columnwidth]{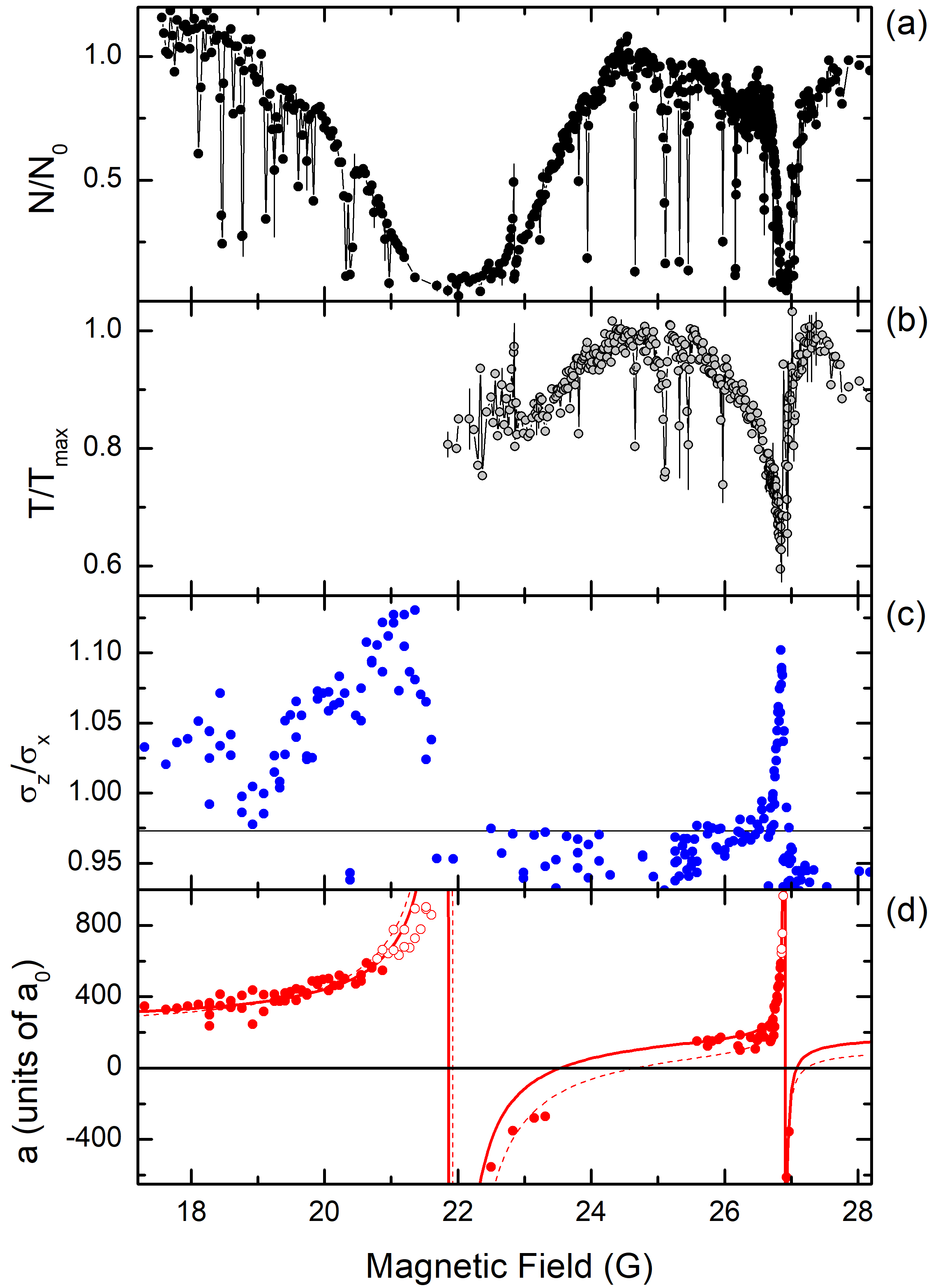}
\caption{(a) High resolution atom loss spectroscopy. Line is a guide to eye. (b) Normalized temperature in the zero-crossing regions after a thermalization experiment. (c) Aspect ratio (AR) of the thermal atomic cloud after 12~ms free expansion from a trap with vertical frequency of 169~Hz and horizontal frequencies of 38~Hz and 107~Hz. (d) Scattering lengths extracted from the data in (c). The dashed and continuous lines are fits of the AR data alone and of the combined AR and binding-energy data, respectively. Open dots are data excluded from the fits. See text for details.}  \label{fig2}
\end{figure}

We now describe the measurements and analysis of Feshbach resonances. We explore the magnetic field range 0-30~G with high resolution (3~mG) performing, as a first step, loss spectroscopy. For this measurement we prepare a thermal sample of about $1.5 \times 10^5$ atoms at a typical temperature of 200~nK by performing evaporation in the crossed dipole trap at B=3.335(3)~G. We then change the magnetic field to the desired value in less than 10~ms and we record the atom number after a waiting time of a few hundreds of ms. For increasing scattering lengths we expect larger loss rates because of enhanced three-body recombination processes. Evaporation ramps and waiting times are slightly adjusted in different subranges of magnetic field in order to optimize the visibility of the Feshbach resonances. In panel (a) of Fig.~\ref{fig2} we show the results of our measurements. Across the entire explored range, we observe the chaotic Feshbach spectrum typical of ultracold magnetic Lanthanide atoms, characterized by many narrow Feshbach resonances with typical widths of 10~mG or smaller and spacing of 100~mG \cite{Frisch2014,Maier2015}. However, around 22~G and 27~G we observe also clear signatures of two broader resonances.

As is well known, across a Feshbach resonance, the scattering length depends on the magnetic field according to
$a(B)=a_{bg}(1-\Delta/(B-B_0))$, where $a_{bg}$ is the background value of the scattering length and $B_0$ and $\Delta$ are the resonance center and width, respectively \cite{Chinrev}. A rough indication for $\Delta$ can be extracted from thermalization measurements, along the lines of previous studies \cite{Kadau2016,Ferlaino,PfauJulienne}. We set the desired magnetic field value before starting the evaporation in the crossed optical trap and we record the temperature after the evaporation ramps (panel (b) of Fig.\ref{fig2}). The idea is that the evaporation efficiency depends on the elastic scattering length: small values of $a$ lead to a poor evaporation efficiency resulting in higher final temperatures. The maximum temperature is therefore expected to be close to the zero-crossing of $a$ and the shift between this point and the dip in the loss spectroscopy is a measurement for $\Delta$. However, a non-negligible contribution of the dipolar interaction, which depends on the sample geometry, might affect the thermalization process, shifting the temperature maximum from the zero of the contact interaction. We indeed observe that the magnetic field delivering the maximum temperature depends the trap frequencies. Therefore, such analysis gives just a rough estimate of the resonances widths: $\Delta_1\approx$ 3~G and $\Delta_2\approx$ 0.3~G.

For a more precise characterization of the two resonances, we employ the complementary technique introduced in ref. \cite{Tang2016}, relying on the anisotropic expansion of a thermal dipolar gas released from the trapping potential. The observable is the aspect ratio (AR) of the atomic sample after a free expansion. The AR is indeed predicted to depend in a known way on the scattering length and on other parameters (trap frequencies, atom number, temperature, magnetic moment, time of flight) \cite{Tang2016}. We perform this measurement at approximately the same temperature as the loss spectroscopy, employing an imaging beam that propagates horizontally. After evaporation, we shift the field to the desired value with a 10 ms linear ramp and we wait for 40 ms before releasing the atoms for TOF imaging. From the measured AR (panel (c) in Fig. \ref{fig2}) we reconstruct the scattering length as a function of the magnetic field (panel (d) in Fig.\ref{fig2}). We estimate a systematic error of $\pm1$\% in the AR measurement, mainly due to inhomogeneity of the imaging beam. This results in  systematic uncertainty on the estimated scattering length that is larger than the statistical fluctuation. Furthermore, we note that for low values of the scattering length, a small variation in the aspect ratio reflects in a large variation in $a$ in particular for low atom number. For this reasons, similarly to ref. \cite{Tang2016}, we have a blind magnetic field region in between the two resonances, which for our parameters corresponds to AR$<$0.975, for which we cannot extract the scattering length in a reliable way. We exclude from the analysis also the regions of very large scattering length where the large loss rates might invalidate the assumption of thermal equilibrium for the model to apply \cite{Tang2016}; we arbitrarily choose to exclude the data with $|a|>600\,a_0$.

We fit the experimental data with the expression for neighboring resonances \cite{overlapping} \be
a(B)=a_{bg}(1-\Delta_1/(B-B_{01})-\Delta_2/(B-B_{02}))\,.
\ee
We obtain $a_{bg}=180(50)\,a_0$ for the background scattering length, $B_{01}=21.93(20)$~G and $\Delta_1=2.9(10)$ G for the center and width of the broad resonance, $B_{02}=26.892(7)$~G and $\Delta_2=0.14(2)$~G for the narrow one. The quoted uncertainties are set by the systematic uncertainty on the AR mentioned above.

\begin{figure}[t]
\includegraphics[width=\columnwidth]{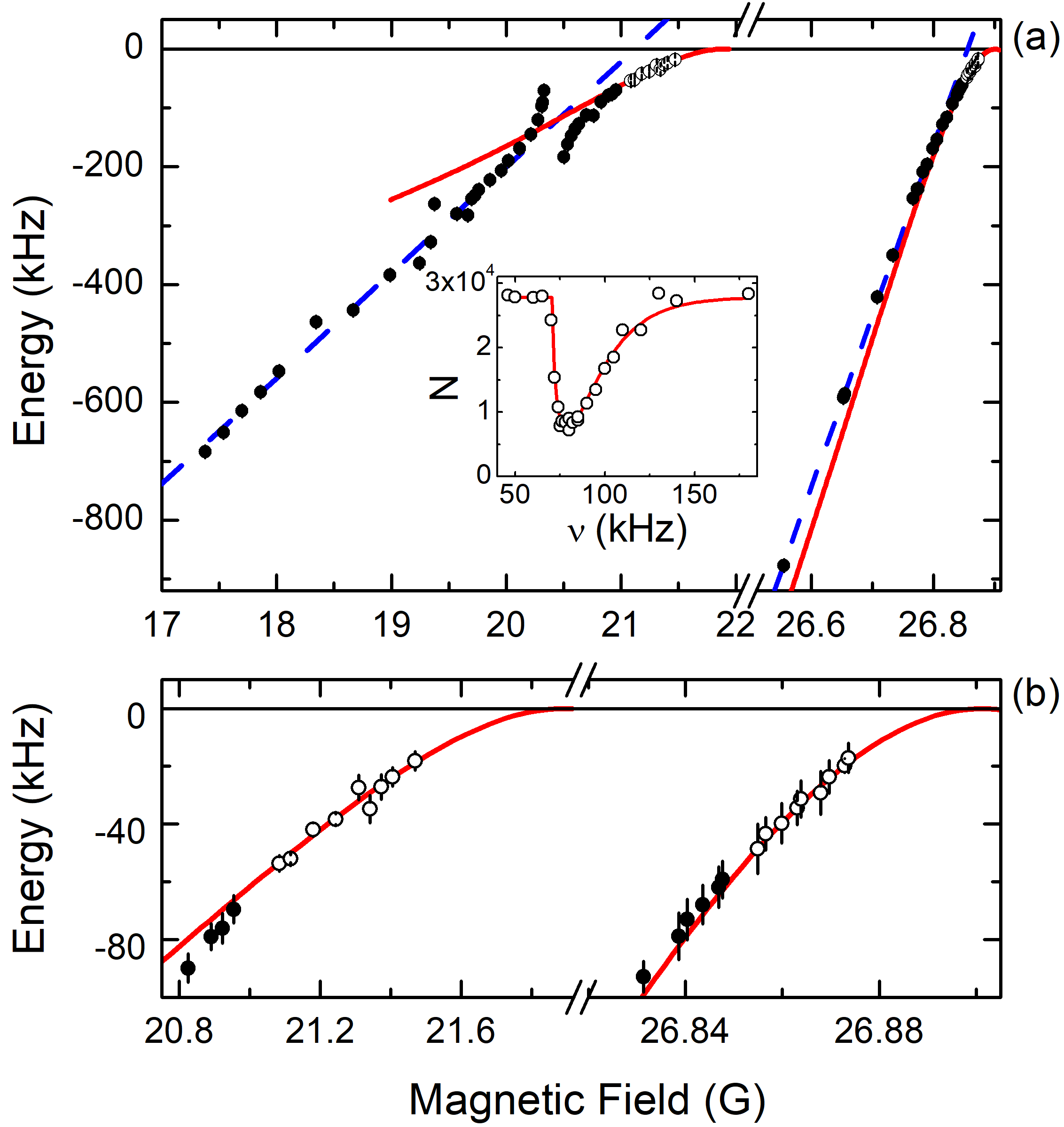}
\caption{Molecular binding energy for the two broad Feshbach resonances, measured by magnetic-field modulation spectroscopy. (a) Full data range. (b) Zoom of the data close to the resonance centers. The continuous red lines are combined fits of the data close to the centers (open dots) and of the AR data with the models of eqs.1-2, see text. The dashed blue lines are linear fits to estimate the magnetic moment of the molecular state. Inset: typical molecular association spectrum. The solid line is a fit with the line shape described in \cite{Jones1999}.}  \label{fig3}
\end{figure}

A further characterization of the Feshbach resonances comes from the measurement of the binding energy of the corresponding molecular states \cite{Chinrev}. We apply a sinusoidal modulation on the magnetic field, with typical peak to peak amplitude of 100~mG, for 200-400~ms. When the modulation frequency matches the binding energy, two atoms are associated into a weakly bound dimer that rapidly decays, leading to atom losses. For each value of the static $B$ field we observe an asymmetric shape of the atomic loss peak profile, which we fit as described in \cite{Jones1999,Napolitano1994}. In Fig. \ref{fig3} we plot the binding energy $E(B)$ for both resonances, measured as the peak position of the fitted loss feature as a function of the magnetic field. For both resonances we can identify a quadratic regime close to $B_0$, and a linear regime far from the resonance centers. For the broad resonance, we also observe several avoided crossings with other molecular states associated to narrow resonances.

As discussed in \cite{Kotochigova}, Feshbach resonances in Lanthanides cannot be typically associated to a single partial wave. For the aim of the present work we attempt basic fits of $E(B)$ with the theoretical models for s-wave resonances. We fit simultaneously the data close to both resonances, in the range $(B_0-B)/\Delta\leq 0.25$ corresponding to $a\gtrsim750\,a_{0}$, with the corrected universal model \cite{Flambaum}
\be
E(B)=\frac{\hbar^2}{m (a(B)-\bar{a})^2}\,,
\label{univ}
\ee
with $a(B)$ given by eq.(1) and where $\bar{a}$ is the mean scattering length related to the van der Waals length scale by $\bar{a}=0.956~R_{\mathrm{vdW}}$. For Dy, $R_{\mathrm{vdW}}=77~a_0$ \cite{PfauJulienne,Kotochigova2011}. For an isolated resonance, such type of fit would determine quite precisely the resonance center $B_0$ and the product $a_{bg}\Delta$. The two resonances we are exploring are instead coupled by eq.(1), in the sense that the narrow resonance experiences a local background scattering length determined by the broad resonance. Therefore, from the fit we can determine reliably only the resonance centers, $B_{01}=22.0(4)$~G and $B_{02}=26.91(2)$~G. These values are consistent with those obtained by the AR measurements.

In order to verify that both the expansion data and the binding energy data can be described by the same model for $a(B)$, we perform a combined fit of the two datasets. We obtain $a_{bg}= 220(50)\,a_0$ for the background scattering length, $B_{01}=21.91(5)$~G and $\Delta_1=1.9(7)$ G for the center and width of the broad resonance, $B_{02}=26.902(4)$~G and $\Delta_2=0.14(5)$~G for the narrow one. The uncertainties are dominated by the statistical uncertainty of the combined fit. These parameters should be more accurate than those determined by the expansion measurements alone, since they are derived from two different observables and therefore are less prone to systematic errors. The nominal behavior of $a(B)$ shown in Fig. \ref{fig2}d (continuous line) differs from that obtained by the fit of the AR data alone (dashed line) mainly for the width of the broadest resonance. Both fits do not reproduce the excluded data close to the resonances, confirming that the model of \cite{Tang2016} might not work for too large scattering lengths. The results of the fit do not change if the cutoff for the AR data is changed in the range 400-800 $a_0$. We note that our background scattering length differs with the value $a_{bg}=157(4)\,a_0$ determined around 5G in ref. \cite{Tang2016}.

From the binding energy measurements we can also extract the magnetic moment of the associated molecules, by performing a linear fit of the binding energy far from the resonance center (blue dash-dotted lines in Fig.\ref{fig3}. In fact, if the resonance is closed-channel-dominated, the linear coefficient of the fit represents the difference between the molecular and the atomic magnetic moments $\delta\mu$. The fit yields $\delta\mu_1$=0.128(5)~$\mu_B$ and $\delta\mu_2$=2.07(2)~$\mu_B$ for the broad and narrow resonance, respectively. The broad resonance is therefore associated to a molecular state that has almost the same magnetic moment as the unbound atoms. On the line of ref. \cite{PfauJulienne}, we estimate the resonance strength using the expression $s_{\mathrm{res}}=a_{bg}\Delta\delta\mu/\bar{a}\bar{E}$, where $\bar{E}$ is related to the van der Waals energy scale by $\bar{E}=1.094~E_{\mathrm{vdW}}$ (for Dy, $E_{\mathrm{vdW}}/h=1.877$~MHz \cite{PfauJulienne,Kotochigova2011}). For the narrow resonance we employ a "local" background scattering length of 170 $a_0$ given by the broad resonance. We obtain the same resonance strength, $s_{\mathrm{res}}=0.5(3)$, for both resonances, despite the different magnetic field widths. The small $s_{\mathrm{res}}$ value suggests two closed-channel-dominated resonances, justifying our analysis. It will be interesting to see whether these rather isolated resonance can be modeled theoretically, perhaps also to confirm a predominant s-wave nature \cite{Kotochigova,PfauJulienne}.

In conclusion, we reported the efficient production of a dBEC of $^{162}$Dy atoms thanks to a resonator-enhanced optical trap, and a characterization of  the scattering properties up to 30~G. The presence of two relatively broad Feshbach resonances, sided by just a few other narrow ones, is interesting in view of a precise tuning of the contact interaction in a wide range of values. In particular, from the loss spectrum in Fig.\ref{fig2} it is possible to note that the magnetic field region around the narrower resonance centered at $B_{02}\simeq 27$ G is promising to access large values of $a$, both positive or negative, and also a region with $a\simeq 0$, with very few interfering resonances. The dipolar BEC with tunable scattering length can be employed for investigating a range of phenomena where the relative dipolar and contact interaction strengths need to be controlled precisely, including those requiring large scattering lengths like the Efimov effect \cite{Greene}.

We acknowledge support by L.A. Gizzi, the EC-H2020 research and innovation program (Grant 641122 - QUIC), the ERC (Grants 203479 - QUPOL and 247371 - DISQUA) and the italian MIUR (PRIN2015 2015C5SEJJ). We also acknowledge discussions with A. Simoni and technical assistance by A. Barbini, F. Pardini, M. Tagliaferri and M. Voliani.



\end{document}